\documentclass[a4paper,12pt]{article}
\usepackage{latexsym}
\usepackage{amsmath}
\usepackage{amssymb}
\usepackage{amscd}

\newtheorem{theo}{Theorem}[section]
\newtheorem{lemm}[theo]{Lemma}
\newtheorem{coro}[theo]{Corollary}
\newtheorem{prop}[theo]{Proposition}
\newtheorem{defi}[theo]{Definition}

\newcommand{\EQ}{\begin{equation}}
\newcommand{\EN}{\end{equation}}
\newcommand{\pr}{\indent{\em Proof.  }}
\newcommand{\qed}{\hspace*{5 mm}$\Box$}
\newcommand{\w}{\mbox{wt}}
\newcommand{\Aut}{\mbox{\rm Aut}}

\newcommand{\bh}{{\bf h}}

\newcommand{\by}{{\bf y}}
\newcommand{\bx}{{\bf x}}

\newcommand{\bv}{{\bf v}}
\newcommand{\bu}{{\bf u}}

\newcommand{\bo}{{\bf 0}}

\newcommand{\F}{\mathbb{F}}

\title{On linear $q$-ary completely regular codes with $\rho
= 2$ and dual antipodal\footnote{This work has been partially
supported by the Spanish MICINN grant MTM2009-08435, the Catalan
AGAUR grants 2009SGR1224, 2009PIV00071 and also by the Russian fund
of fundamental researches, project no. 09 -01 - 00536.}}

\author{J. Borges, J. Rif\`{a} \\
Department of Information and Communications Engineering,\\
 Universitat Aut\`{o}noma de Barcelona,\\
  08193-Bellaterra, Spain\\
 (emails:~\{joaquim.borges,josep.rifa\}@autonoma.edu)\\
\and V. A. Zinoviev\\Institute for Problems of Information
Transmission,\\ Russian Academy of Sciences,\\ Bol'shoi Karetnyi
per. 19, GSP-4, Moscow,\\ 127994, Russia (e-mail:\, zinov@iitp.ru).}

\begin{document}

\maketitle

\begin{abstract}
We characterize all linear $q$-ary completely regular codes with
covering radius $\rho=2$ when the dual codes are antipodal.  These
completely regular codes are extensions of linear completely regular
codes with covering radius 1, which are all classified. For
$\rho=2$, we give a list of all such codes known to us. This also
gives the characterization of two weight linear antipodal codes.
\end{abstract}

\section{Introduction and Terminology}

Let $\F_q=GF(q)$ be the Galois field with $q$ elements, where $q$ is
a prime power. $\F^n_q$ is the $n$-dimensional vector space over
$\F_q$. Let $\w(\bv)$ be the {\em Hamming weight} of a vector $\bv
\in \F_q$, and $d(\bv, \bu)=\w(\bv-\bu)$ denotes the {\em Hamming
distance} between two vectors $\bv,\bu \in \F_q^n$. Say that vectors
$\bv$ and $\bu$ are {\em neighbors} if $d(\bv,\bu)=1$.

A $q$-ary {\em code} $C$ of length $n$ is a subset of $\F_q^n$. If
$C$ is a $k$-dimensional linear subspace of $\F^n_q$, then $C$ is a
$q$-ary {\em linear} code, denoted by $[n,k,d]_q$, where $d$ is the
{\em minimum distance} between any pair of codewords, assuming
$k\geq 1$. If any pair of distinct codewords are at distance $d$,
then the code $C$ is {\em equidistant}. A linear code $C$ will be
called \textit{antipodal} if it contains a vector of weight equal to
the length of the code. Say that a linear $[n,k,d]_q$ code is {\em
trivial}, if $k\leq 1$ or $k \geq n-1$.

For a $q$-ary code $C$ with minimum distance $d$ denote by
$e = \lfloor (d-1)/2 \rfloor$ its {\em packing radius}.
Given any vector $\bv \in \F^n_q$, its {\em distance} to the code
$C$ is
\[
d(\bv,C)=\min_{\bx \in C}\{ d(\bv, \bx)\}
\]
and the {\em covering radius} of the code $C$ is
\[
\rho=\max_{\bv \in \F^n_q} \{d(\bv, C)\}.
\]
Clearly $e\leq\rho$ and $C$ is {\em perfect} if and only if
$e=\rho$. It is well known that the only nontrivial linear perfect
codes for $e=\rho=1$ are the {\em Hamming codes}, which have length
$n=(q^m-1)/(q-1)$, dimension $k=n-m$ and minimum distance $3$. We
denote by $H_m$ a parity check matrix of a Hamming code, that is,
any vector $\bv$ is a codeword if and only if $H_m\cdot
\bv^t=\bo^t$, where $\bo^t$ is the all-zero column vector.

For any $\bx\in \F^n_q$, let $~D=C+\bx~$ be a {\em coset} of  $C$. A
{\em leader} of $D$ is a minimum weight vector in $D$.

For a given $q$-ary code $C$ with covering radius
$\rho=\rho(C)$ define
\[
C(i)~=~\{\bx \in \F_q^n:\;d(\bx,C)=i\},\;\;i=1,2,...,\rho.
\]

\begin{defi}\label{de:1.1}  A code $C$ is completely regular,
if for all $l\geq 0$ every vector $x \in C(l)$ has the same number
$c_l$ of neighbors in $C(l-1)$ and the same number $b_l$ of
neighbors in $C(l+1)$. Also, define $a_l = (q-1){\cdot}n-b_l-c_l$ and note
that $c_0=b_\rho=0$. The intersection  array of $C$ is $(b_0,
\ldots, b_{\rho-1}; c_1,\ldots, c_{\rho})$.
\end{defi}

%\begin{defi}\label{de:1.2}
%A partition of $\F^n_q$ into disjoint subsets
%\[
%\F_q^n = \bigcup_{i=0}^{\rho} \E_i
%\]
%is called {\em equitable}, or, equivalently, {\em perfect
%$(\rho+1)$-coloring}, with quotient matrix $[\mu_{i,j}]$,~$i,j
%\in \{0,\ldots,\rho\}$, if for all $i$ and $j$, any vector in
%$\E_i$ has exactly $\mu_{i,j}$ neighbors in $\E_j$.
%\end{defi}

%Given a completely regular code $C \subset \F_q^n$ with covering
%radius $\rho$, the partition $\F_q^n = C \cup C(1) \cup \ldots \cup
%C(\rho)$ is clearly equitable \cite{bro2,gods}. But, in general, the
%inverse is not true. In the specific case when $\rho=1$ the inverse
%is true and, hence, in this case an equitable partition is
%equivalent to a completely regular code.

A {\em linear automorphism} of $\F^n_q$ is a coordinate permutation
together with a product by a nonzero scalar value at each position.
Such an automorphism $\sigma$ can be represented by a $n\times n$
monomial matrix $M$ such that $\bx M=\sigma(\bx)$, for all
$\bx\in\F^n_q$. Two codes, $C$ and $C'$, are {\em equivalent} if
there is a linear automorphism of $\F^n_q$, say $\sigma$, such that
$C'=\sigma(C)$. From now on, if $C\subseteq \F^n_q$ is a linear
code, the {\em full automorphism group} of $C$, denoted $\Aut(C)$,
is the group of linear automorphisms of $\F^n_q$ that leaves $C$
invariant. We say that $\Aut(C)$ is {\em transitive} if all one
weight vectors in $\F^n_q$ are in the same orbit. For a linear code
$C$, the group $\Aut(C)$ acts on the set of cosets of $C$ in the
following way: for all $\phi \in \Aut(C)$ and for every vector $\bv
\in \F^n_q$ we have $\phi(\bv + C) = \phi(\bv) + C$.

\begin{defi}[\cite{sole,giud}]\label{de:1.3}
Let $C$ be a linear $q$-ary code with covering radius $\rho$. Then
$C$ is completely transitive if $\Aut(C)$ has $\rho + 1$ orbits when
acts on the cosets of $C$.
\end{defi}

Since two cosets in the same orbit should have the same weight
distribution, it is clear that any completely transitive code is
completely regular.

\begin{lemm}[\cite{sole}]\label{TransitiveCT}
Let $C$ be a $[n,k,d]_q$ code with covering radius $\rho=1$. If
$\Aut(C)$ is transitive, then $C$ is completely transitive.
\end{lemm}

\pr Obvious, since all cosets of $C$, different of $C$, have leaders
of weight $1$. Thus, all such cosets are in the same orbit. \qed

The next known result follows from Singer theorem~\cite{sing}.

\begin{lemm}\label{HammingTrans}
Let ${\mathcal H}$ be a $[n,k,3]_q$ Hamming code. Then,
$\Aut({\mathcal H})$ is transitive. Hence, any such code is
completely transitive.
\end{lemm}

\medskip

It has been conjectured~\cite{neum} for a long time that if $C$ is a
completely regular code and $|C|>2$, then $e \leq 3$. For the case
of binary linear completely transitive codes, the problem of
existence is solved: it is proven in \cite{bor1,bor2} that for $e
\geq 4$ such nontrivial codes do not exist. The conjecture is also
proven for the case of perfect codes ($e=\rho$) \cite{tiet,zino} and
quasi-perfect ($e+1=\rho$) uniformly packed codes \cite{sem2,tilb}
(defined and studied also in \cite{bas1,Lits,goet}).

When $e\leq 3$, there are many well known completely regular codes
and, recently, we have presented new constructions of binary and
non-binary completely regular codes \cite{bor3,rif1,rif2}. However,
there does not exist a general classification of completely regular
codes with $e\leq 3$. Since $d\in\{2e+1,2e+2\}$, $e\leq\rho$ and any
perfect code has odd $d$, we have that the minimum distance of a
completely regular code with $\rho=1$ is $d\leq 3$ and for the codes
with $\rho=2$ is $d\leq 5$. In this paper we classify all linear
$q$-ary completely regular codes with $\rho=1$ and we also
characterize the structure of linear completely regular codes with
$\rho = 2$, whose dual is antipodal. We also list all such codes we
know.

After submission of \cite{bor4} we found \cite{kool}, where a large
class of so called {\em arithmetic completely regular} codes has
been classified. In particular, in \cite{kool}, it also appears the
classification of all linear $q$-ary completely regular codes with
$\rho = 1$. The approach in \cite{kool} is based on known results on
classification of distance regular graphs in Hamming schemes. Our
approach here is self contained and based only on classical results
on perfect and uniformly packed codes. Both approaches are
interesting from the point of view of classification of completely
regular codes with small covering radius $\rho$, in particular, for
enumeration of all completely regular codes with small parameters
$q\,n \leq 48$, suggested by Neumaier \cite{neum}. We emphasize
that, unlike the linear case, the classification of nonlinear
completely regular codes with $\rho=1$ as well as of nonlinear
equidistant codes are hard problems (see, for example,
\cite{fdf1,fdf2} for the first object and \cite{bogd}, and
references there, for the second).

The paper is organized as follows. In Section 2 we construct for any
prime power $q$ and any integers $n$ and $k$, such that $n \geq q+1$
and $2 \leq k \leq n-2$, all nontrivial, nonequivalent, linear
$q$-ary completely regular $[n,k,d]_q$ codes with covering radius
$\rho=1$. This also gives the construction of all the linear $q$-ary
equidistant codes. We prove that the constructed codes are the only
linear $q$-ary completely regular codes with $\rho=1$ and show that
all such completely regular codes are completely transitive. In
Section 3 we consider linear $q$-ary nontrivial completely regular
$[n,k,d]_q$ codes with covering radius $\rho=2$, whose dual is
antipodal. We give necessary and sufficient conditions for the shape
of the parity check matrices of such codes known to us and we give
their intersection arrays. Also, we point out some remarkable
properties (e.g. self-duality) and equivalence to the existence of
some two weight codes.

\section{Classification of linear $q$-ary completely regular codes with
$\rho=1$}

We consider nontrivial linear  $[n,k,d]_q$ codes,
i.e. $k$ is in the region $2 \leq k \leq n-2$.

\begin{defi}\label{de:2.1}
Let $C$ be a $q$-ary code of  length $n$ and let $\rho$ be its
covering radius. We say that $C$ is {\em uniformly packed} in the
wide sense, i.e. in the sense of \cite{bas1}, if there exist
rational numbers $~\beta_0,\ldots,\beta_{\rho}~$ such that for any
$\bv\in\F_q^n$ \EQ\label{eq:2.1}
\sum_{k=0}^{\rho}\beta_k\,\alpha_k(\bv)~=~1, \EN where
$\alpha_k(\bv)$ is the number of codewords at distance $k$ from
$\bv$.
\end{defi}

Note that the case $\rho=e+1$ and $\beta_{\rho-1} = \beta_{\rho}$
corresponds to {\em strongly uniformly packed} codes \cite{sem2} and
the case $\rho=e+1$ corresponds to {\em uniformly packed} codes
\cite{goet}.

For a $[n,k,d]_q$ code $C$ let $(\eta^{\perp}_0, \ldots,
\eta^{\perp}_n)$ be the weight distribution of its dual code
$C^{\perp}$, assume $(\eta^{\perp}_0, \ldots, \eta^{\perp}_n)$ has
$s = s(C)$ nonzero components $\eta^{\perp}_i$ for $1 \leq i \leq
n$. Following to Delsarte \cite{dels}, we call $s$ the {\em external
distance} of $C$.

\begin{lemm}\label{lem:2.2}
Let $C$ be a code with minimum distance $d$, packing radius
$e=\lfloor \frac{d-1}{2} \rfloor$, covering radius
$\rho$ and external distance $s$. Then:\\
(i) \cite{dels} $\rho \leq s$.\\
(ii) \cite{bas2} $\rho = s$ if and only if $C$ is uniformly
packed in the wide sense.\\
(iii) \cite{bro2} If $C$ is completely regular, it is uniformly
packed in the wide sense.\\
(iv) \cite{goet, sem2} If $C$ is uniformly packed in the wide sense
and $\rho=e+1$, then it is completely regular. \end{lemm}

The next fact follows from Lemma \ref{lem:2.2}. Earlier, it was
mentioned for $q=2$ in \cite{mac1} and for $q \geq 2$ (for example,
in \cite{sem1}).

\begin{lemm}\label{lem:2.4} Let $C$ be a nontrivial linear $q$-ary code with
$d = 3$ and let $C^\perp$ be its dual. Then $C$ is a Hamming code,
if and only if $C^\perp$ is equidistant.
\end{lemm}

The next statement is a simple generalization of Lemma \ref{lem:2.4}
to the case of arbitrary completely regular codes with $\rho=1$.

\begin{lemm}\label{lem:2.5} Let $C$ be a nontrivial linear $q$-ary
code and $C^\perp$ be its dual. Then $C$ is completely regular with
$\rho=1$ if and only if $C^\perp$ is equidistant.
\end{lemm}

So, to classify all linear completely regular codes with $\rho=1$ we
have to classify all linear equidistant codes. It has been done in
\cite{boni} when a code does not contain trivial (zero) positions,
but this is not enough for completely regular codes with $\rho=1$.

%We use the notation ${\mathcal H}$ for a Hamming $[n_m,k,3]_q$ code
%with parity check matrix $H_m$, where $n_m = (q^m-1)/(q-1)$ and $k =
%n_m - m$.

\begin{defi} $($Construction $I(u))$.
Let $C$ be a $[n,k,d]_q$ code with a parity check matrix $H$. Define
a new code $C^{+u}$ with parameters $[n+u,k+u,1]_q$ as the code with
parity check matrix $H^{+u}$, obtained by adding $u>0$ zero columns
to $H$.
\end{defi}

The following statement follows directly from the definition of
$C^{+1}$.

\begin{lemm}\label{construction1}
Let $C$ be a $[n,k,d]_q$ code and let $C^{+1}$ be obtained from $C$
by Construction $I(1)$. Let $\bx=(x_1,\ldots,x_n)\in\F^n_q$ be at
distance $i$ apart from $\alpha_i$ codewords in $C$ and at distance
$i-1$ apart from $\alpha_{i-1}$ codewords in $C$. Then, for any
$x_{n+1} \in \F_q$ the vector $\bx'=(x_1,\ldots,x_n\,|\,x_{n+1})$ is
at distance $i$ apart from exactly $\alpha_i+(q-1)\alpha_{i-1}$
codewords in $C^{+1}$.
\end{lemm}

\begin{prop}\label{prop:2.1}
Codes $C$ and $C^{+u}$ have the same covering radius and, moreover,
$C$ is completely regular if and only if $C^{+u}$ is completely
regular. In this case, both codes have the same intersection
numbers, i.e.
\[
a'_i~=~a_i+(q-1)u,~~b'_i~=~b_i,~~c'_i~=~c_i,~~i =0,1,\ldots,\rho.
\]
\end{prop}

\pr
It is enough to consider the case $u=1$. For any vector
$\bx=(x_1,\ldots,x_n)\in \F^n_q$ denote by
$\bx'=(x_1,\ldots,x_n\,|\,x_{n+1})$ the corresponding $q$ vectors
from $\F^{n+1}_q$. Let $\by=(y_1,\ldots,y_n)\in C$ be a codeword at
distance $\rho$ from $\bx$. Then $\by'=(y_1,\ldots,y_n\,|\,x_{n+1})$
is a codeword in $C^{+1}$ at the same distance $\rho$ from $\bx'$.
Therefore, $C$ and $C^{+1}$ have the same covering radius $\rho$.

Assume $C$ is completely regular. For any vector
$\bx=(x_1,\ldots,x_n)\in \F_q^n$ at distance $t\leq \rho$ from $C$,
denote by $\alpha_{i,t}$ the number of codewords in $C$ at distance
$i$ from $\bx$ $(0\leq i \leq n)$. As $C$ is completely regular,
$\alpha_{i,t}$ does not depend on $\bx$, but just on $t$
and $i$. Take a vector $\bx'=(x_1,\ldots,x_n\,|\,x_{n+1}) \in
\F_q^{n+1}$, which is at distance $t$ from $C^{+1}$. It is easy to
see that the number of codewords in $C^{+1}$ at distance $i$, say
$\alpha_{i,t}'$, depends only on $t$ and $i$. Indeed, by Lemma
\ref{construction1} we have
$\alpha_{i,t}'=\alpha_{i,t}+(q-1)\alpha_{i-1,t}$, for all
$i=0,\ldots,n$, and $\alpha_{n+1,t}'=(q-1)\alpha_{n,t}$.

Conversely, assume that $C$ is not completely regular. Let
$\bx,\by\in \F_q^n$ be such that $d(\bx,C)=d(\by,C)=t>0$ and, for
$0\leq i\leq n$, let $\alpha_{i,t}(\bx)$ (respectively,
$\alpha_{i,t}(\by)$) denote the number of codewords at distance $i$
from $\bx$ (respectively, from $\by$). Since $C$ is not completely
regular, we can select $\bx$ and $\by$ such that
$\alpha_{i,t}(\bx)\neq \alpha_{i,t}(\by)$ for some $i\geq t$. Let
$i$ be the minimum possible of such values, that is
$\alpha_{i-1,t}(\bx)=\alpha_{i-1,t}(\by)$. Then, by Lemma
\ref{construction1}, for the corresponding vectors $\bx'$ and
$\by'$, we have $\alpha_{i,t}'(\bx')\neq \alpha_{i,t}'(\by')$.
Consequently, $C^{+1}$ is not completely regular. \qed

As a summary, starting with any completely regular code, we
obtain an infinite family of completely regular codes with the
same covering radius.

\begin{defi} (Construction $II(\ell)$).
Let $C$ be a $[n,k,d]_q$ code with parity check matrix $H$. Let
$C^{\times \ell}$ be the code with parameters
$[n\,\ell,k+(\ell+1)n,2]_q$, whose parity check matrix, denoted
$H^{\times \ell}$, is $\ell$ times the repetition of $H$ (or nonzero
multiples of $H$), i.e.
$$
H^{\times \ell}~=~[H^{(1)}\,|\,H^{(2)}\,|\,\cdots \,|\,H^{(\ell)}],
$$
where $H^{(i)}$ is scalar nonzero multiple of $H$, for all
$i=1,\ldots,\ell$.
\end{defi}

%As we said before, $n_m = \frac{q^m-1}{q-1}$ is the length of a
%perfect $q$-ary (Hamming) $[n_m,k=n_m-m,3]_q$ code ${\mathcal H}$,
%with parity check matrix $H_m$, where $m \geq 2$ and hence $n_m \geq
%q+1$.

\begin{prop}\label{prop:2.2}
A $[n,k,d]_q$ code $C$ is completely regular with covering radius
$\rho=1$ if and only if $C^{\times \ell}$ is completely regular with
covering radius $\rho'=1$.
%\begin{enumerate}
%\item[(i)] If $H=H_m$ and $\ell \geq 2$, then $d'=2$.
%\item[(ii)] If $H=H_m^{+u}$, $u>0$ and $\ell \geq 2$, then $d'=1$.
%\item[(iii)] If $d=2$ and $\ell \geq 2$, then $d'=2$.
%\end{enumerate}
\end{prop}

\pr If $C$ is completely regular with $\rho=1$, its parity check
matrix $H$ is the generator matrix of an equidistant code since, by
Lemma \ref{lem:2.2}, the external distance $s$ equals to $\rho=1$.
The matrix $H^{\times \ell}$ generates such a code too. Hence,
$C^{\times \ell}$ has an external distance $s=1$ and, by Lemma
\ref{lem:2.2}, the covering radius $\rho'=1$. We deduce, again by
Lemma \ref{lem:2.2}, that $C^{\times \ell}$ is completely regular.
The converse statement follows by using the same arguments, if we
take into account the shape of the matrix $H^{\times \ell}$. \qed

Finally, we summarize the main results of this section.

\begin{theo}\label{theo:3.1}
Let $C$ be a nontrivial $[n,k,d]_q$ code with covering radius
$\rho=1$. Then, $C$ is completely regular if and only if its parity
check matrix is of the form
\[
H~=~\left((H_m)^{\times \ell}\right)^{+u},
\]
(up to column permutations), where $H_m$ is the parity check matrix
of a Hamming code of length $n_m=(q^m-1)/(q-1)$. The length and
dimension of $C$ are $n=n_m\,\ell+u$ and $k=n-m$, respectively.

Furthermore
\begin{itemize}
\item[(i)] $d=3$, if and only if $u=0$, $\ell = 1$, $n = n_{m}$
and $C$ is a Hamming code.
\item[(ii)] $d=2$, if and only if $u=0$, $\ell \geq 2$, $n =
n_{m}\ell$.
\item[(iii)] $d=1$, if and only if $u>0$, $\ell \geq 1$.
\item[(iv)] The code $C$ has the intersection numbers:
\[
a_0=(q-1)u,\;\; b_0=(q-1)\ell\, n_{n-k},\;\; c_1=\ell,\;\;
a_1=(\ell\,n_{n-k}+u)(q-1)-\ell.
\]
\item[(v)] The code $C$ is completely transitive.
\end{itemize}
\end{theo}

\pr
The ``if part" is clear combining Constructions $I(u)$ and
$II(\ell)$.

For the ``only if part", since $\rho=1$ we deduce that
$d\in\{1,2,3\}$.

We separate these three cases:

(i) If $d=1$, then $H$ has zero columns. Thus, using Construction
$I(u)$, $C$ can be obtained from a completely regular code with
minimum distance greater than 1 and covering radius 1.

(ii) If $d=2$, since $2 \leq k \leq n-2$ and $\rho=1$, the matrix
$H$ generates the equidistant $[n,n-k,d^{\perp}]_q$ code
$C^{\perp}$. As $d=2$ this matrix $H$ contains repeated columns and
does not contain zero columns.

First, we prove that every column of $H$ occurs the same number, say
$\ell$ times, where $\ell \geq 2$ (counting includes, of course, the
columns, obtained multiplying by scalar elements from $\F_q$).
Assume that each column $\bh$ occurs $\ell_{\bh}$ times. Since $C$
is completely regular the intersection number $c_1$ is the same for
any vector $\bx \in C(1)$. Take such a vector of weight $1$ with its
nonzero position at the column $\bh$. Then $c_1(\bh)$ is equal to
$\ell_{\bh}+1$ (we take into account the zero codeword). We conclude
that the number $\ell_{\bh}$ should be the same for every column in
the matrix $H$, i.e. $\ell_{\bh} = \ell$.

The last equality implies that $n$ should be divisible by $\ell$.
Denote $n' = n/\ell$. Present the matrix $H$ in the form
\[
H~=~[H'| \dots |H'],
\]
where all columns in $H'$ are different.

Now, we claim that $n' = n_m=(q^m-1)/(q-1)$ for some $m \geq 2$,
i.e. the matrix $H'$ is the parity check matrix of a Hamming
$[n_m,n_m-m,3]_q$ code and so, the matrix $H$ generates an
equidistant code. Since each row of $H$ is $\ell$ times the
repetition of the same vector, the matrix $H'$ also generates an
equidistant code, say $E$ of length $n'$. Since the dual code of $E$
has minimum distance $d \geq 3$, covering radius $\rho = 1$ and
external distance $s=1$ (Lemma \ref{lem:2.2}; it is a perfect code
of the length $n_m = (q^m-1)/(q-1)$ for some $m \geq 2$ (Lemma
\ref{lem:2.4}). If $m=1$, then $H$ consists of only one row,
implying that $d'=2$.

Finally, we conclude that $k = n - m$ and $C$ is obtained by
Construction $II(\ell)$ from the perfect (Hamming) $[n_m,n_m-m,3]_q$
code.

(iii) If $d=3$, since $\rho=1$, $C$ is a perfect code by definition.

(iv) It is straightforward to find the intersection numbers.

(v) If $d\in\{2,3\}$, using Lemma \ref{HammingTrans}, $C$ is
equivalent to a code $C'$ such that $\Aut(C')$ is transitive. Thus,
$\Aut(C)$ is transitive and, by Lemma \ref{TransitiveCT}, $C$ is
completely transitive.

If $d=1$, then let $D$ be the `reduced' code, that is, the code
obtained from $C$ by doing the reverse operation of Construction
$I(u)$. Since both, the covering radius of $C$ and $D$ are 1, we
have that $C\neq \F^n_q$ and, by Proposition \ref{prop:2.1}, $D$ is
a completely regular code with $d>1$. Hence, $D$ is a completely
transitive code. This means that we can choose a set of $q^{n-k}-1$
coset leaders of weight one such that they are in the same orbit of
$\Aut(D)$. But $C$ and $D$ have the same number of cosets and we can
choose, in both $C$ and $D$, the same coset leaders. Since
$\Aut(D)\subseteq \Aut(C)$, we have that these coset leaders are in
the same orbit. Therefore, all the cosets different of $C$ are in
the same orbit and $C$ is a completely transitive code. \qed

The next statement follows directly from Theorem \ref{theo:3.1} and
it has been obtained in~\cite{boni}, for the case of codes without
trivial (zero) positions.

\begin{coro}\label{theo:5.1}
Given an equidistant $[n,k,d]_q$ code we have that $n =
n_{n-k}\,\ell + u$ for some $\ell \geq 1$ and some $u \geq 0$.
Furthermore, a generator matrix is obtained from $H_{n-k}$ (a parity
check matrix of the Hamming code) by repeating this matrix $\ell$
times and then adding $u$ trivial (zero) columns.
\end{coro}

\section{Linear $q$-ary completely regular codes with
$\rho=2$}

In this section we deal with linear $q$-ary nontrivial completely
regular $[n, k, d]_q$ codes with covering radius $\rho=2$, whose
dual is antipodal and we show  a characterization of these codes
using the ones with covering radius one.

Now, we recall some facts on extension of codes. For a $q$-ary code
$C$ of length $n$, the extended code $C^*$ of length $n+1$ is
obtained by adding an overall parity check position. This means that
for any codeword $\bx=(x_1,\ldots,x_{n+1})\in C^*$, we have
$\sum_{i=1}^{n+1} x_i=0$ (where the sum is in $\F_q$). We say that
such an extension {\em works} if $d^*=d+1$, where $d^*$ is the
minimum distance in $C^*$ and $d$ is the minimum distance in $C$.
Generally speaking, extension of equivalent codes can result in
different codes, but if an extension works for two equivalent codes,
then the resulting codes would have the same parameters. The
following result is well known and can be found, for example, in
\cite{mac2}.

\begin{lemm}\label{lem:exten}
Let ${\mathcal H}_m$ be a $[n_m,k,3]_q$ Hamming code. Then, the
extended code ${\mathcal H}_m^*$ has minimum distance 4 if
\begin{enumerate}
\item[(i)] $q=2$ and $m \geq 2$, or
\item[(ii)] $q=2^r \geq 4$ and $m=2$, i.e. $n_m+1 = q + 2$ and
$k=q-1$.
\end{enumerate}
\end{lemm}

Denote by $H_m^*$ the parity check matrix of the $[n_m+1, k, 4]$
code, obtained as the extended Hamming code ${\mathcal H}_m$ for a
case when extension works. In this case, denote such code by
${\mathcal H}^*_m$. Denote by $D_m$ the matrix of size $(m+1) \times
q^m$ whose columns are all the $q^m$ vectors of length $m \geq 1$
with an extra $(m+1)$th position equal to $1$; (this matrix, without
the $(m+1)$th row, generates a {\em difference matrix} \cite{comb};
this is why we denote it by $D_m$). We remark that this matrix $D_m$
can also be obtained by repeating the $q-1$ different multiples of
the matrix $H_m$ and by adding, first, a zero column and, finally,
the all-one row of length $q^m$.

We recall a result in \cite{del1}. For a given $[n,k,d]_q$ code $C$
with parity check matrix $H$ define its {\em complementary}
$[n_{n-k} -n,k,\bar{d}]$ code $\bar{C}$, whose parity check matrix
$\bar{H}$ is obtained from the matrix $H_{n-k}$ by removing all the
columns of $H$ and multiples of them. Recall an important property
of complementary codes: {\em to any codeword of weight $w$ in a
$[n,k,d]_q$ code $C$ corresponds a codeword of weight
$\bar{w}=q^{n-k-1} - w$ in the complementary code $\bar{C}$}. As a
corollary of this fact above we have the next lemma.

\begin{lemm}\label{lem:complementary} \cite{del1}
A linear projective $[n,k,d]_q$ code $C$ with covering radius $\rho
= 2$, which is not a difference-matrix code, does exist
simultaneously with its complementary projective code $\bar{C}$ with
the same covering radius $\bar{\rho} = 2$.
\end{lemm}

\begin{theo}\label{theo:4.1}
Let we have a nontrivial $[n,k,d]_q$ code $C$. Let $H$ be its parity
check matrix. Then, $C$ is completely regular with covering radius
$\rho=2$ and the dual code $C^{\perp}$ is antipodal if and only if
the matrix $H$ looks,up to equivalence, as follows:
\[
H=\left[
\begin{array}{cccc}
\,1 \,&\,\cdots \,&\,1\,\\
\, &\,\,M\,\,&\,\,\\
\end{array}
\right],
\]
where $M$ generates an equidistant code $E$ with the following
property: for any nonzero codeword $\bv\in E$, every symbol
$\alpha\in\F_q$, which occurs in a coordinate position of $\bv$,
occurs in this codeword exactly $n-\tilde{d}$ times, where
$\tilde{d}$ is the minimum distance of $E$. Moreover, up to
equivalence, $C$ is the extension of a completely regular code $C'$
with covering radius $\rho'=1$.
\end{theo}

\pr
The code $C^{\perp}$ is antipodal if and only if $H$ contains the
all-one row, up to equivalence. Now, by Lemma \ref{lem:2.2}, $C$ is
completely regular with $\rho=2$ if and only if the external
distance is $s=2$. Equivalently, $H$ generates a code, the dual code
$C^\perp$, with two different weights and $M$ generates an
equidistant code $E$ such that every symbol $\alpha\in\F_q$, which
occurs in a coordinate position of a codeword $\bv\in E$, occurs in
this codeword exactly $n-\tilde{d}$ times, where $\tilde{d}$ is the
minimum distance of $E$. Notice that if $M$ has this property, we
can add any multiple of the all-one row to any row of $M$ and we do
not change this property.

Also, up to equivalence, we can rewrite $H$ in the following form:

\[
\left[
\begin{array}{cccc}
\,1 \,&\,\cdots \,&\,1\,&\,1\,\\
\, &\,\,H'\,\,&\,\,&\,\bf{0}\,\\
\end{array}
\right],
\]
where $\bo$ is the zero column of length $n-k-1$. Hence, $H'$
generates words of only one weight. This means that $H'$ is a parity
check matrix for a $[n-1,k,d']_q$ code $C'$ (which is obtained by
puncturing the last coordinate of $C$) with external distance $s'=1$
and, by Lemma \ref{lem:2.2}, covering radius $\rho'=1$. Therefore,
$C'$ is a completely regular code and it must be one of the cases of
Theorem \ref{theo:3.1}.

We conclude that any nontrivial completely regular code with
covering radius $2$, whose dual code is antipodal, is obtained from
some code with covering radius $1$ by adding the overall parity
checking position. \qed

The following statement is a direct corollary of Theorem
\ref{theo:4.1}.

\begin{coro}\label{theo:5.2}
Let we have a nontrivial two-weight $[n,k,d]_q$ code $C$ with
weights $w_1$ and $w_2 = d$ and with generator matrix $G$. If
$w_1=n$, then
\[
G=\left[
\begin{array}{cccc}
\,1 \,&\,\cdots \,&\,1\,&\,1\,\\
\, &\,\,M\,\,&\,\,&\,\bf{0}\,\\
\end{array}
\right],
\]
where $M$ generates an equidistant $[n-1,k-1,d]_q$ code $E$ with the
following property: for every codeword $\bv\in E$, every symbol
$\alpha\in\F_q$ which occurs in $\bv$, occurs in $\bv$ exactly $n-d$
times.
\end{coro}

Now, we enumerate the completely regular codes with $\rho=2$, whose
dual is antipodal, that we know. We also compute the intersection
array for all the enumerated codes. Some of these codes were
mentioned in \cite{goet}. Dual of these codes are {\em two-weight}
antipodal codes mostly due to Delsarte \cite{del1}, studied by many
other authors (see a nice survey of two-weight codes in
\cite{cald}). We do not know if the list is exhaustive but any other
such code, according to Theorem \ref{theo:4.1}, would also be an
extended completely regular code with $\rho=1$ .

\begin{prop}\label{enum}
The following codes are completely regular with
covering radius $\rho=2$ and their dual codes are antipodal.

\begin{enumerate}

\item[(i)] The binary extended perfect $[n,k,4]_2$
code ${\mathcal H}^*_{m}$ of length $n = 2^m$, where $k=n-m-1$ and
$m \geq 2$. Its intersection array is
$$
(n,n-1;1,n).
$$

\item[(ii)] The extended perfect $[n,k,4]_q$
code ${\mathcal H}^*_{m}$ of length $n=q+2$ with $k=q-1$, where
$q=2^r \geq 4$, and $m=2$ \cite{bush,del1} (the family $TF1$ in \cite{cald}). Its intersection array
is
$$
\left((q+2)(q-1),q^2-1;1,q+2\right).
$$

\item[(iii)] The difference-matrix $[n,k,3]_q$ code of length
$n = q^m$, dimension $k = n - (m+1)$ with parity check matrix $D_m$,
where $m \geq 1$, and $q \geq 3$ is any prime power (the dual code
generated by the matrix $D_m$ has been given in \cite{sem1}). The
complementary code of this code is the Hamming code ${\mathcal
H}_{m}$ and its intersection array is
$$
\left(n(q-1),n-1;1,n(q-1)\right).
$$

\item[(iv)] Latin-square $[n,n-2,3]_q$ code of length
$n$,  with parity check matrix $H$, obtained from $D_1$ by deleting
any $q-n$ columns, where $3 \leq n \leq q$ and $q \geq 3$ is any
prime power \cite{del1}. Its intersection array is
$$
\left(n(q-1),(q-n+1)(n-1);1,n(n-1)\right).
$$

\item[(v)] A $[n=q(q-1)/2,k=n-3,4]_q$ code
for $q=2^r \geq 4$ \cite{del1} (the complementary code belongs to
$TF1^d$, i.e. it is the projective dual code to the code (ii)
(family $TF1$ in \cite{cald}). Its intersection array is
$$
\left((q-1)n,(q-2)(q+1)(q+2)/4; 1, q(q-1)(q-2)/4\right).
$$

\item[(vi)] A $[n = 1 + (q+1)(h-1),k=n-3,4]_q$ code, where
$1 < h < q$ and $h$ divides $q$, for $q=2^r \geq 4$ (the family
$TF2$ in \cite{cald}). Its intersection array is
$$
\left((q-1)n,(q+1)(h-1)(q-h+1); 1,(h-1)n\right).
$$

\item[(vii)] A $[n = q(q-h+1)/h,k=n-3,4]_q$ code, where
$1 < h < q$ and $h$ divides $q$, for $q=2^r \geq 4$ (the
complementary belongs to the family $TF2^d$ \cite{cald}). Its
intersection array is
$$
\left((q-1)n,(q+1)(q-h)(q(h-1)+h)/h^2; 1, q(q-h)(q-h+1)/h^2 \right).
$$

\end{enumerate}
\end{prop}

Cases (i) and (ii) correspond to extended codes of case (i) in
Theorem \ref{theo:3.1}.  Case (iii) corresponds to an extended code
$C$ of case (ii) in Theorem \ref{theo:3.1}, where $\ell=q-1$ and the
parity check matrix $H$ of $C$ is
$$
H~=~[H_m^{(1)}| \dots |H_m^{(q-1)}],
$$
where each $H_m^{(i)}$ is a different scalar multiple of $H_m$. Case
(iv) corresponds to an extended trivial completely regular code of
co-dimension $1$ and covering radius $1$. Finally, cases (v) - (vii)
correspond to extension of several (multiple) copies of completely
regular codes of co-dimension $2$ and covering radius $1$.

Notice, we can apply Construction $II(\ell)$ to any of the codes
above, obtaining a completely regular code with $\rho=2$ and $d=2$.
Also, we can apply Construction $I(u)$ to anyone of these codes,
including those obtained by Construction $II(\ell)$, obtaining a
code with $\rho=2$ and $d=1$.

To finish this section it is proper to emphasize one interesting
class of codes, which belong to the family (iv).

Let $C$ be a $q$-ary linear code of length $n$ and parity check
matrix $H$. For any integer $r\geq 1$, we define the {\em lifted
code} \cite{rif3} $C'\subset \F_{q^r}^n$ as the linear code which
has parity check matrix $H$. Two nice properties of lifted codes are
the following.

\begin{lemm}\label{lemm:self}
Let $C$ be a $[n,k,d]_q$ code and let $C'\subset\F_{q^r}^n$ be any
lifted code. Then, $C$ is self-dual ($C=C^\perp$) if and only if
$C'$ is self-dual.
\end{lemm}

\pr Note that $C$ is self-dual if and only if $n=2k$ and the rows of
its parity check matrix $H$ are orthogonal. But $n$, $k$ and $H$ do
not change for $C'$ and, since $\F_q^n$ is a subspace of
$\F_{q^r}^n$, the rows of $H$ are orthogonal vectors in $\F_{q^r}^n$
if and only if they are orthogonal in $\F_q^n$. \qed

\begin{prop}\label{selfduality}
Let $C'\subset\F_{q^r}^n$ be a lifted code from a Hamming
$[n,n-m,3]_q$ perfect code ${\mathcal H}_m$. Then $C'$ is self-dual
if and only if ${\mathcal H}_m$ is a ternary $[4,2,3]_3$ Hamming
code.
\end{prop}

\pr
If $C'$ is self-dual, then $|C'|=|(C')^\perp|$ implying that
$n-m=m$, i.e. $n=2m$. But the length of a $q$-ary Hamming code is
$$
n=\frac{q^m-1}{q-1}=\sum_{i=1}^m q^{m-i}.
$$
Hence, we immediately obtain that $m=2$, $q=3$ and $n=4$. A parity
check matrix for a Hamming $[4,2,3]_3$ ${\mathcal H_2}$ code is

\[
H_2=\left[
\begin{array}{cccc}
\,0 \,&\,1\,&\,1\,&\,1\,\\
\,1 \,&\,0\,&\,1\,&\,2\,\\
\end{array}
\right].
\]
Since the rows of $H_2$ are orthogonal, ${\mathcal H_2}$ is a
self-dual code. Finally, by Lemma \ref{lemm:self}, if $C'$ is a
lifted code from ${\mathcal H_2}$, then $C'$ is self-dual. \qed

Notice that the lifted perfect $[q+1,q-1,3]_{q^r}$ codes (see
\cite{rif3}), with parity check matrix $H_2$ over $\F_q$  are
particular cases of the family (iv) in Proposition \ref{enum}, for
$r>1$. So, these codes are completely regular with intersection
array
$$
\left((q+1)(q^r-1),q^2(q^{r-1}-1);1,q(q+1)\right).
$$
According to Proposition \ref{selfduality}, the case $q=3$
corresponds to a self-dual code, for any $r$. Actually, self-dual
codes with $\rho=2$ exist for any prime power $q\geq 4$.

\begin{prop}\label{prop:dual}
Let $q \geq 4$ be any prime power and let $\F_q = \{0,1,\xi_2,
\ldots, \xi_{q-1}\}$. Let the matrix $D^{q-4}_1$,
\[
D^{q-4}_1~=~\left[
\begin{array}{ccll}
~1~&~1~&~1~&~1\\
~0~&~1~&~\xi_i~&~\xi_j\\
\end{array}
\right],
\]
be a parity check matrix for the code $C$ and a generator matrix for
the code $C^{\perp}$, where  $\xi_i, \xi_j \in \F^*_q$ are two different elements such that $\xi_i~+~\xi_j~+~1~=~0$. Then
$C$, as well as $C^{\perp}$, is a linear antipodal completely
regular $[4,2,3]_q$ code with covering radius $\rho = 2$ and with
intersection array $(4\,(q-1), 3\,(q-3); 1, 12)$. Furthermore, for
the case $q=2^r \geq 4$, these two equivalent codes coincide: $C =
C^{\perp}$, i.e. $C$ is self-dual.
\end{prop}

\pr It is straightforward that, when $q=2^r\geq 4$, the equation
$\xi^2_i~+~\xi^2_j~+~1~=~0$ is always satisfied when
$\xi_i~+~\xi_j~+~1~=~0$. \qed

We notice that there are nonlinear completely regular $q$-ary
Latin-square codes with length $n=4$, cardinality $q^2$, minimum
distance $d=3$; covering radius $\rho=2$ and with the same
intersection array $((q-1)\,4, 3\,(q-3); 1, 12)$ that the codes in
Proposition \ref{prop:dual}. The existence is guaranteed for any
integer $q \geq 3$ with one exception for $q = 6$ \cite{comb}; since
two orthogonal Latin squares of order $6$ do not exist.

\end{document}